\documentclass[graybox]{svmult}

\usepackage{type1cm}        
\usepackage{makeidx}         
\usepackage{graphicx}        
\usepackage{multicol}        
\usepackage[bottom]{footmisc}
\usepackage{todonotes}
\usepackage{hyperref}
\usepackage{url}

\usepackage{newtxtext}       %
\usepackage{newtxmath}       

\makeindex             

\begin{document}

\title*{Teaching Empirical Research Methods in Software Engineering: An Editorial Introduction}
\author{Daniel Mendez ,\\ 
Paris Avgeriou, \\ 
Marcos Kalinowski, \\
Nauman bin Ali}
\institute{Daniel Mendez \at Blekinge Institute of Technology and
fortiss \email{daniel.mendez@bth.se} \and Paris Avgeriou \at University of Groningen \email{p.avgeriou@rug.nl} \and Marcos Kalinowski \at Pontifical Catholic University of Rio de Janeiro, \email{kalinowski@inf.puc-rio.br} \and Nauman bin Ali \at Blekinge Institute of Technology \email{nauman.ali@bth.se}}

\maketitle


\section{Introduction}
\label{sec:1}

Empirical software engineering has received much attention in recent years and marked a certain shift from a more mathematically-oriented and design-science-driven engineering discipline to a more insight-oriented and theory-driven one~\cite{mendez19EMSEInterdiscipline}. Before that, a large extent of the available body of knowledge in Software Engineering was rooted in rationalist arguments where ``common knowledge'' was established on the sole ground of reasoning by argument (and, in some cases, reasoning by authority), logical inference, and mathematical proofs. While this may still be the case for certain areas within Software Engineering, our discipline as a whole has evolved and grown significantly in terms of scientific maturity; so has the need for excellence in empiricism, as we now reason, in simple terms, by evidence gained through sensory experiences\footnote{We borrow the term ``sensory experiences'' from the established terminology in the philosophy of science~\cite{Stanley14} and subsequently use it in the widest possible sense. As software artefacts and phenomena can materialise in various ways, the term may well include direct and indirect observations (e.g., via measurements).} (e.g., observations) -- while following certain rules, principles, and norms for scientific practice~\cite{Stanley14}.

The ambition is that decisions in software engineering practice become rooted in empirical evidence \cite{Kitchenham+2004,Ali16}. Those decisions may well range from the choice of software engineering methods, modelling and description techniques, and tools, over to testing the sensitivity of the software-intensive products and services themselves to the characteristics and needs of their practical contexts (i.e. fit for purpose). The importance of empirical software engineering research\footnote{In the remainder of this chapter, we refer with \emph{empirical software engineering} to the research activities aiming at producing evidence (i.e., conducting and reporting empirical studies) rather than to the act of consuming evidence.} is, therefore, well recognised by both researchers and practitioners, and competencies in empirical software engineering are in high demand. 

It is therefore not surprising that more empirically oriented forums like the Empirical Software Engineering journal (ESE), the International Symposium on Empirical Software Engineering and Measurement (ESEM) and the International Software Engineering Research Network (ISERN), or the International Conference on Evaluation and Assessment in Software Engineering (EASE) have become part of the prestigious venues in the software engineering research community \cite{WongMAL21}. Moreover, even the more traditional, leading venues like IEEE Transactions on Software Engineering (TSE) or the International Conference on Software Engineering (ICSE) have come to expect an empirical evaluation of proposed ideas as a de-facto prerequisite to merit the inclusion of manuscripts in their programs.

To become a reflective software engineer \cite{DybaMG14}, practitioners need to be equipped with the abilities and skills for critical thinking and reasoning, systematic problem-solving, and objective decision-making not only based on reasoning by (rationalist) arguments, but also based on empirical insights and data. In fact, leveraging empirical software engineering competencies is a pre-requisite for continuous improvement~\cite{Basili1994ExperienceFactory,dittrich2020exploring} as they enable organisations to test new ideas, processes, and technologies, fostering a learning environment where they can adapt and evolve based on empirical data~\cite{Basili1994ExperienceFactory,dittrich2020exploring}. High software development maturity has consequently been related to establishing empirical and data-driven continuous improvement capacities to ensure that innovations align with the business goals and contribute to enhanced quality and process performance~\cite{CMMIV2.0, kalinowski2014results}.

The demand for competencies in empirical software engineering gave rise to various courses on empirical software engineering in both higher education and professional training programmes. The goals of these courses range from preparing students for their graduate programmes and thesis projects, to providing them with essential skills for a career in software engineering research and practice.

While extensive guidelines are available for designing, conducting, reporting, and reviewing empirical studies, similar attention has not yet been paid to teaching empirical software engineering. We, therefore, recognise that there is a need to gather experiences, best practices, and lessons learned from the community to jointly develop and inform each other on possibilities for effectively teaching the principles and methods of empirical software engineering. Nowadays, courses on empirical software engineering are mostly designed independently (thus, re-inventing the wheel every time and sometimes proposing competing or even conflicting concepts and ideas) without relying on a shared basis for knowledge, materials, and what could be considered good practices.

Closing this gap sets the aim and scope of this edited book. Before introducing the scope of the book, we attempt to motivate the challenges in Empirical Software Engineering through the lenses of educators. This, in turn, requires a basic understanding of what Empirical Software Engineering is, which we address next.

\section{Empirical Software Engineering in a Nutshell}
\label{sec: EMSE}
As Empirical Software Engineering may come with different interpretations, also due to the lack of a commonly used definition, we introduce the very basic concepts and terms next.

Software engineering research and practice are continuously evolving to cope with the emerging challenges imposed by the ubiquitous nature and rapid advancements in engineering methods and technologies used. This poses various challenges for the selections and configurations of engineering teams, methods, and tools that can no longer be answered by rationalist arguments only.

In response to the emerging challenges in software engineering, we have already observed a prominent turn towards empirical approaches. On the one hand, the recognition of the importance of empirical research to software engineering allows us scholars to critically reflect upon central and, to some extent, more philosophical questions like \emph{What qualifies as scientifically sound practice?} or \emph{What are relevant research questions and appropriate research methods to address a certain research problem?}. On the other hand, it has also yielded various more practical contributions that impact software engineering practice. 

The community of researchers has already made strong contributions that allowed us to shift from a purely design science-oriented discipline driven by the application of scientific concepts and methods to practical ends -- rendering the discipline as a more traditional \emph{engineering} discipline -- to a more insight-oriented and theory-driven (\emph{scientific}) discipline~\cite{WOHLIN2015229}. 

This shift towards a more insight-oriented discipline based on empirical evidence has taken place in recent years with a growing and vibrant community of researchers coining it as \emph{Empirical Software Engineering (Research)} (EMSE)~\cite{Kitchenham+2004, Rombach2013, Basili:2013:Perspectives}. 

In simple terms and for the sake of this brief introduction, the ultimate goal of Empirical Software Engineering is to advance our body of knowledge by continuously building and evaluating scientific theories.

\begin{shaded}\textbf{Scientific Knowledge and Theories}

With \emph{Scientific Knowledge}, we refer to a portrait of our (human) understanding of reality which is captured via scientific theories. 

With \emph{Scientific Theory}, we refer to the belief that there are patterns in phenomena while having survived (1) tests against sensory experiences and (2) criticism by critical peers. With the first criterion, we refer to what is commonly known as the demarcation problem introduced by Karl Popper~\cite{popper1959logic}. Here, the main argument is that a theory qualifies only as a scientific theory (rather than a non-scientific or a pseudo-scientific one), if the underlying hypotheses are testable and, most importantly, falsifiable. With the second criterion, we refer to quality assurance of scientific practices and safeguards implemented by the research community, such as peer review. This may of course render the acceptance of scientific results also as a social process.

\end{shaded}

Empirical Software Engineering has relevance to both academic research communities and software engineering practice alike, as it allows us, eventually, to:
\begin{enumerate}
    \item Reason about our discipline and the involved phenomena (e.g., social phenomena).
    \item Recognise and understand the limitations and effects of artefacts in their practical contexts (e.g., by evaluating technologies, techniques, processes, or models).
\end{enumerate}

\begin{figure}[!ht]
\centering
\includegraphics[width=\columnwidth]{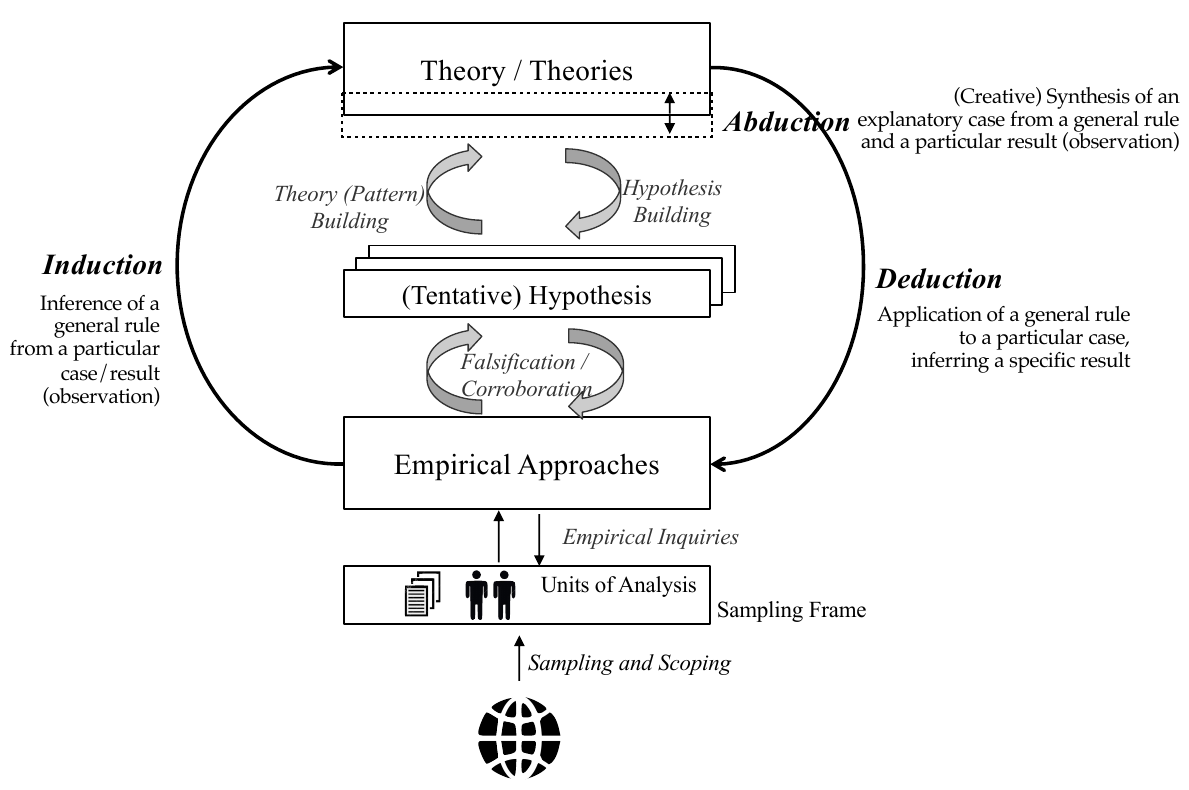}
\caption{Cycle for empirical research, taken from ~\cite{mendez19EMSEInterdiscipline}.}
\label{fig:empiricism_nutshell}
\end{figure}

To picture a process, if there is any at all, that could characterise the general way of working in Empirical Software Engineering, we believe that it is best described by what we introduced as a pragmatic cycle, as shown in~\ref{fig:empiricism_nutshell} and taken from~\cite{mendez19EMSEInterdiscipline}. That cycle characterises in its essence the different perspectives we may take on the notion of theories in Software Engineering and consequently the approaches we may take towards building and evaluating theories:

\begin{itemize}
\item \emph{Induction} describes the inference of a general rule from a particular case and is often referred to as the principle of building/extending theories, for example, by conducting observational studies to explore practical, real-life contexts and relevant parameters.
\item \emph{Deduction} describes the application of a general rule to a particular case and is often referred to as theory evaluation, where we test whether the hypothesised outcome of a theory can also be observed in reality; for example, by conducting quasi-controlled experiments to corroborate concepts by testing the expected improvements.
\item \emph{Abduction} describes the, to some extent, creative synthesis of a case from a general rule and a particular result and used, for example, to infer probable cases out of existing theories. 
\end{itemize}

In Empirical Software Engineering, assuming an ideal scenario, we would follow that cycle multiple times to contextualise and increase the robustness of our theories and, eventually, scale them up to practice by triangulating through different settings and methods (both quantitative and qualitative ones).

While we refer to~\cite{mendez19EMSEInterdiscipline, felderer2020evolution} for a more profound discussion of Empirical Software Engineering and its historical evolution, we may conclude here, for the sake of our introductory chapter, that every empirical approach and every epistemological stance~\footnote{With epistemology, we refer to the theory of knowledge.}, of which there are many, has its well-deserved place in that larger picture that has been elaborated under the assumption of the following postulates:

\begin{enumerate}
    \item There is nothing absolute about truth in the notion of scientific theories, but good (commonly accepted) scientific principles and norms to build and evaluate theories and, thus, increase our confidence (level of evidence). 
    \item There is no such thing as a universally valid way of working; in fact, there is no rigid, let alone standardised process but many different and valid (and complementary) ways of undertaking research.
    \item There is no single, empirically inquired point of view that will ever provide us with an entire picture when interpreting relevant real-world phenomena. 
\end{enumerate}

As a consequence of the above, elaborating strong, robust theories in the community -- that is, theories with a high level of confidence -- also requires multiple iterations through the cycle (e.g. with replication studies) using complementary research methods while constantly setting the result of each inquiry in relation to existing evidence.

We therefore recognise well that there is no purity in Empirical Software Engineering and no single empirical approach that would be best for all purposes. At the same time, we have still reached a common understanding in the community of researchers of the basic concepts and terms, and we can currently observe already excellent contributions to research and practice. This is also corroborated by edited collections in our community that contribute to laying out more fundamental positionings, experience-based recommendations and guidelines, and general approaches to conduct Empirical Software Engineering research, such as~\cite{felderer2020contemporary}. These are further complemented by attempts to standardising quality criteria of reported studies~\cite{ralph2021acm}.

The question that remains now open is how to educate and train students and software practitioners in those concepts. More specifically, there are certain challenges in doing so, as elaborated in the following section.

\section{Challenges when Teaching Empirical Software Engineering}

When teaching Empirical Software Engineering, we face no fewer challenges than when engaging in empiricism in research and practice. We believe that among the most pressing challenges that educators face are the following three:

\textbf{Lack of purity.} Software Engineering is to some extent also a social practice~\cite{dittrich2016does} and as such it incorporates also the same plurality as the social sciences. In that sense, there is no purity in Empirical Software Engineering research, same as there is no purity in Software Engineering practice. Software Engineering research and practice are characterised by many different families of systems, domains, subject areas, and disciplinary backgrounds, many coming from natural sciences. This has effects on preferred principles and understanding, often coming with some forms of prejudice (e.g., artificially separating qualitative from quantitative camps). As no empirical approach fits best for all purposes, students should be taught how to choose or combine suitable methods based on research goals and context and adapt them based on reflective evaluations of outcomes. Embracing this lack of a standardised way of working is, therefore, a fundamental virtue to empirical research in Software Engineering, but this makes teaching it inherently challenging. With chapters taking a broader perspective on the variety of research methods and their combination, such as the chapter on research designs in Software Engineering, we attempt to (partially) address this challenge.

\textbf{Teaching EMSE requires more than teaching the details of single research methods.} While we may well agree on basic concepts and terminology, at least in the long run, we will still be confronted with the limitations given in classroom settings where students will have difficulties experiencing and understanding the core of Empirical Software Engineering (research). This is also because Empirical Software Engineering goes well beyond the scope of a single more traditional research discipline in, for example, natural sciences. It requires a more profound understanding of its philosophical roots and its practical implications. For example, how to appropriately frame a research problem, how to identify relevant research questions, what relevant means both from a theoretical and a practical perspective, and how to choose and apply research methods and adequate data analysis techniques makes clear that teaching EMSE goes beyond teaching single research methods, but requires a profound understanding of a larger picture where philosophical arguments and trade-offs in the choice of research methods are the norm. This is especially true when scaling method application and analysis. We try to address this challenge, at least to some extent, with chapters focusing more on the fundamentals of empirical software engineering, such as on instrumentation and measurements, or on theorising in Software Engineering research.

\textbf{Difficulty in teaching flexible design studies.} Teaching flexible design research is additionally hard in traditional courses because: (a) the design of the study evolves based on the findings, (b) it is inconceivable even to foresee all design decisions that will need to be taken, (c) it often involves qualitative data collection and analysis. These characteristics make it difficult to experience such research in a course. On the contrary, for a fixed design study involving an experiment or a questionnaire-based survey with quantitative data, the students can experience the challenges of designing a complete study and, through pilots, even experience operationalising and reporting the study. With several chapters on teaching flexible design studies like action research, case studies, and ethnographies, this book aims to support educators in teaching these particularly challenging methods.

The difficulty we faced when planning this book project was, therefore, to balance the need of recognising the above outlined diversity in Empirical Software Engineering while still preserving some common ground on which we all can agree.

\section{Scope of the Book}

This book draws from the profound experience and expertise of well-established scholars in the Empirical Software Engineering community to provide guidance and support in teaching the various approaches and philosophical stances. 

A particular focus is on combining (1) research methods and their epistemological settings and terminology, and the (2) didactics and pedagogy for the subject. The book covers the most essential and contemporary research methods and philosophical and cross-cutting concerns in software engineering research, considering both academic and industrial settings. It also provides insights into the effective teaching of concepts and strategies. 

Rather than providing a set of prescriptive blueprints for blindly setting up courses, our ambition was to provide the readers with both a more generic (customisable) basis for course design and practical advice and support; the latter comes in the form of experience reports, best practices, annotated bibliographies/readers, slide decks, datasets, available tools, prerequisite knowledge, examples/exercises (formative), and assignments --  found in the subsequent chapters and the online material sections. This can equip educators with the necessary information and tools for effectively preparing students for today’s prevalent needs in software engineering research and practice.

\begin{shaded}\textbf{Online Repository for this Edited Book}

Many of the chapters refer to online material including, if not limited to, bibliographies/readers, slide decks, datasets, tools, examples/exercises, or assignments. This repository is centrally available through \url{www.emse.education } and may be updated regularly. A permanently archived, synchronised, and citable Zenodo release can be found under \url{https://zenodo.org/doi/10.5281/zenodo.11544897}.

\end{shaded}

This edited book at hand is therefore meant to (1) provide a common foundation for educators while (2) recognising (and cherishing) the broad spectrum of possible flavours. We hope to have achieved this by approaching this edited book as a joint community project right from its conception. The idea for this book was born as part of a community workshop conducted at the International Software Engineering Research Network (ISERN) and discussed further with members of the research community before yielding an open call for contributions. The editorial and peer-review process was further facilitated by that community with frequent exchanges, feedback, and collegial support. 

We take special pride in the outcome that constitutes a collection of topics covering the broad spectrum of Empirical Software Engineering we believe is important to be represented in educational settings and written by a variety of internationally well established scholars.

In this book, we distinguish essentially two major categories of chapters:
\begin{enumerate}
    \item Chapters reflecting upon more cross-cutting concepts, such as measurement or mining data repositories, as well as chapters offering teaching experience reports. 
    \item Chapters emphasising the sharing of experience-based advice for effectively teaching selected empirical methods. 
\end{enumerate}

Finally, in both categories, we distinguish between more experience-based reflections that may well come with outlines of entire courses and a reflection of lessons learnt, and more foundational chapters that are meant to support readers in setting up their own courses by providing the necessary knowledge and advice. We trust that the reader will find both categories of chapters useful and that they will know how to take both as orientation when designing their own courses.

In the second editorial chapter described next, we introduce a community-inspired syllabus and general reflection upon general contents when teaching empirical research methods. There, we also set the chapter contributions into context to provide an outline of the edited book.

\subsection*{Online Material}

In our online material repository \url{https://zenodo.org/doi/10.5281/zenodo.11544897}, the reader can find different slide decks that introduce the topic area of Empirical Software Engineering in different educational settings, each having with special emphasis on building and evaluating theories. 

\begin{acknowledgement}
We would like to express our profound gratitude to all authors for the various thought-provoking and fruitful exchanges, their feedback and peer-reviews, and their valuable contributions, all making this edited book on Teaching Empirical Research Methods in Software Engineering possible. We would also like to thank Yvonne Dittrich for inspiring exchanges and invaluable feedback on earlier versions of this particular chapter.

\end{acknowledgement}

\bibliographystyle{spmpsci}
\bibliography{literature}
\end{document}